\documentclass{article}

\usepackage[utf8]{inputenc}
\usepackage{amsmath, amsfonts, dsfont}
\usepackage{graphicx}

\usepackage{longtable, booktabs}

\usepackage{clrscode3e}

\usepackage[numbers,square]{natbib}

\usepackage{geometry}
\title{An Evolutionary Optimization Approach to Risk Parity Portfolio Selection}

\author{
Ronald Hochreiter
}

\date{January 2015}

\begin{document}

\maketitle

\begin{abstract}
In this paper we present an evolutionary optimization approach to solve
the risk parity portfolio selection problem. While there exist convex
optimization approaches to solve this problem when long-only portfolios
are considered, the optimization problem becomes non-trivial in the
long-short case. To solve this problem, we propose a genetic algorithm
as well as a local search heuristic. This algorithmic framework is able
to compute solutions successfully. Numerical results using real-world
data substantiate the practicability of the approach presented in this
paper.
\end{abstract}

\section{Introduction}\label{introduction}

The portfolio selection problem is concerned with finding an optimal
portfolio $x$ of assets from a given set of $n$ risky assets out of a
pre-specified asset universe such that the requirements of the
respective investor are met. In general, investors seek to optimize
their portfolio in regard of the trade-off between return and risk, such
that the meta optimization problem can be formulated as shown in Eq.
(\ref{f:meta1}).

\begin{eqnarray}
\begin{array}{ll}
\text{minimize} & \text{\tt Risk}(x) \\
\text{maximize} & \text{\tt Return}(x) \\
\label{f:meta1}
\end{array}
\end{eqnarray}

This bi-criteria optimization problem is commonly reduced to a
single-criteria problem by just focusing on the risk and constraining
the required mean, i.e.~the investor sets a lower expected return target
$\mu$, which is shown in Eq. (\ref{f:meta2}).

\begin{eqnarray}
\begin{array}{ll}
\text{minimize} & \text{\tt Risk}(x) \\
\text{subject to} & \text{\tt Return}(x) \geq \mu \\
\label{f:meta2}
\end{array}
\end{eqnarray}

Markowitz \cite{markowitz1952portfolio} pioneered the idea of
risk-return optimal portfolios using the standard deviation of the
portfolios profit and loss function as risk measure. In this case, the
optimal portfolio $x$ is computed by solving the quadratic optimization
problem shown in Eq. \ref{f:marko1}. The investor needs to estimate a
vector of expected returns $r$ of the assets under consideration as well
as the covariance matrix $\mathbb{C}$. Finally the minimum return target
$\mu$ has to be defined. Any standard quadratic programming solver can
be used to solve this problem numerically.

\begin{eqnarray}
\begin{array}{ll}
\text{minimize} & x^T \mathbb{C} x \\
\text{subject to} & r \times x \geq \mu \\
& \sum x = 1 \\
\label{f:marko1}
\end{array}
\end{eqnarray}

While this formulation has been successfully applied for a long time,
criticism has sparked recently. This is especially due to the problem of
estimating the mean vector. To overcome this problem one seeks
optimization model formulations that solely depend on the covariance
matrix. Sometimes even simpler approaches are favored, e.g.~the 1-over-N
portfolio, which equally weights every asset under consideration. It has
been shown that there are cases, where this simple strategy outperforms
clever optimization strategies, see e.g.~DeMiguel et al.
\cite{demiguel2009optimal}.

Of course, the Markowitz problem can be simplified to a model without
using returns easily by dropping the minimum return constraint. In this
case one receives the Minimum Variance Portfolio (MVP), which is overly
risk-averse.

One important technique used for practical portfolio purposes are
risk-parity portfolios, where the assets are weighted such that they
equally contribute risk to the overall risk of the portfolio. The
properties of such portfolios are discussed by Maillard et al.
\cite{maillard2010erc} and alternative solution approaches are shown by
Chaves et al., see \cite{chaves2011rp} and \cite{chaves2012rp}, as well
as Bai et al. \cite{bai2013rp}.

In this paper, an evolutionary optimization approach to compute optimal
risk parity portfolios will be presented. Evolutionary optimization
approaches have been shown to be useful for solving a wide range of
different portfolio optimization problems, see e.g.
\cite{sharma2012portfolio} or \cite{Hochreiter2008} and the references
therein. See also the series of books on Natural Computing in Finance
for more examples \cite{ncfin2008}, \cite{ncfin2009}, \cite{ncfin2010}.

This paper is organized as follows. Section
\ref{risk-parity-portfolio-selection} describes the risk-parity problem
in detail, Section \ref{implementation} presents the evolutionary
algorithm developed for solving the problem, and Section
\ref{numerical-results} presents numerical results. Finally, Section
\ref{conclusion} concludes the paper.

\section{Risk Parity Portfolio
Selection}\label{risk-parity-portfolio-selection}

The type of risk-parity portfolios discussed in this paper are also
called Equal Risk Contribution (ERC) portfolios. The idea is to find a
portfolio where the assets are weighted such that they equally
contribute risk to the overall risk of the portfolio.

We follow Maillard et al. \cite{maillard2010erc} in their definition of
risk contribution, i.e.~reconsider the above mentioned portfolio
$x = (x_1, x_2, \ldots, x_n)$ of $n$ risky assets. Let $\mathbb{C}$ be
the covariance matrix, $\sigma^2_i$ the variance of asset $i$, and
$\sigma_{ij}$ the covariance between asset $i$ and $j$. Let $\sigma(x)$
be the risk (i.e.~standard deviation) of the portfolio as defined in Eq.
(\ref{f:risk1}).

\begin{equation}
\sigma(x) = \sqrt{x^T \mathbb{C} x} = \sum_i x_i^2 \sigma^2_i + \sum_i \sum_{j \neq i} x_i x_j \sigma_{ij}.
\label{f:risk1}
\end{equation}

Then the marginal risk contributions $\partial_{x_{i}} \sigma(x)$ of
each asset $i$ are defined as follows

\[\partial_{x_{i}} \sigma(x) = \frac{\partial \sigma(x)}{ \partial x_{i} } = \frac{ x_i \sigma^2_i + \sum_{j \neq i} x_j \sigma_{ij} }{\sigma(x) }.\]

If we are considering long-only portfolios then the optimal solution can
be written as an optimization problem containing a logarithmic barrier
term which is shown in Eq. (\ref{f:rp1}) and where $c$ is an arbitrary
positive constant. See e.g.~also \cite{spinu2013rp} for an alternative
formulation. In this long-only case, a singular optimal solution can be
computed.

\begin{eqnarray}
\begin{array}{ll}
\text{minimize} & x^T \mathbb{C} x - c \sum_{i=1}^n \ln x_i \\
\text{subject to} & x_i > 0. \\
\label{f:rp1}
\end{array}
\end{eqnarray}

However, if we want to include short positions then we need to find
solutions in other orthants than in the non-negative orthant. See Bai et
al. \cite{bai2013rp} for a log-barrier approach in this case, which is
shown in Eq. (\ref{f:rp2}).

\begin{eqnarray}
\begin{array}{ll}
\text{minimize} & x^T \mathbb{C} x - c \sum_{i=1}^n \ln \beta_i x_i \\
\text{subject to} & \beta_i x_i > 0, \\
\label{f:rp2}
\end{array}
\end{eqnarray}

where $\beta = (\beta_1, \beta_2, \ldots, \beta_n) \in \{-1, 1\}^n$
defines the orthant where the solution should be computed. For each
choice of $\beta$ the above optimization problem is convex and can be
solved optimally. However, as shown in \cite{bai2013rp} there are $2^n$
different solutions. Investors may add additional constraints to specify
their needs, however this cannot be modeled as one convex optimization
problem, which is why an evolutionary approach is presented here. The
general formulation of the long-short risk parity portfolio problem can
be formulated as Eq. (\ref{f:rp3}) as shown in \cite{maillard2010erc}.

\begin{eqnarray}
\begin{array}{ll}
\text{minimize} & \sum_{i=1, j=1}^n ( x_i(\mathbb{C}x)_i - x_j(\mathbb{C}x)_j )^2 \\
\text{subject to} & a_i \leq x_i \leq b_i, \\
 & \sum_{i=1}^n x_i = 1. \\
\label{f:rp3}
\end{array}
\end{eqnarray}

\section{Implementation}\label{implementation}

The solution is computed in two steps. First, a genetic algorithm will
be employed and afterwards a local search algorithm will be applied.

\subsection{Genetic Algorithm}\label{genetic-algorithm}

We are using a standard genetic algorithm to compute risk-parity optimal
portfolios. The algorithm was implemented using the statistical
computing language R \cite{R2014}.

The fitness definition in the risk-parity setting is given by the
deviance of each risk contribution from the mean of all risk
contributions. Let us use the shorthand notation of
$\Delta_i = \partial_{x_{i}} \sigma(x)$, so we compute the expectation
$\Delta = \mathbb{E}(\Delta_i)$ and define the fitness $f$ as the sum of
the quadratic distance of each risk contribution from the mean. This
non-negative fitness value $f$ has to be minimized, where

\[ f = \sum_i (\Delta_i - \Delta)^2 \]

We use a genotype-phenotype equivalent formulation, i.e.~we use
chromosomes of length $n$ which contain the specific portfolio weights
of the $n$ risky assets. Thus, an important operator is the repair
operator, i.e.~the sum of the portfolio is normalized to $1$ after each
operation.

The genetic operators used in the algorithm can be summarized as
follows:

\paragraph{Elitist selection:}

The best $n_{ES}$ chromosomes of each population are kept in the
population.

\paragraph{Mutation:}

A random selection of $n_{M}$ chromosomes of the parent population will
be mutated. Up to a number of 15\% of the length of the respective
chromosome will be changed to a random value between the portfolio
bounds. Let $\ell$ be the length of the chromosome. First, a random
number between 0 and 0.15 is drawn. This number is multiplied by $\ell$
and rounded up to the next integer value. This value represents the
number of genes to be mutated. The mutation positions will be chosen
randomly. Afterwards the randomly selected positions will be replaced
with a random value between the upper and the lower investment limit of
the respective asset.

\paragraph{Random addition:}

$n_{R}$ new and completely random chromosomes are added to each new
population.

\paragraph{Intermediate crossover:}

Two chromosomes from the parent population will be randomly selected for
an intermediate crossover. The mixing parameter between the two
chromosomes will also be chosen randomly. $n_{IC}$ crossover children
will be added to the next population. Let the mixing parameter be
$\alpha$ and the two randomly chosen parent chromosomes $p_1$ and $p_2$
with genes $p_{1,1}, \ldots, p_{1,\ell}$ and
$p_{2,1}, \ldots, p_{2,\ell}$ respectively, where $\ell$ is the length
of the chromosome. An intermediate crossover will result in a child
chromosome $c$ where the genes are set to
\[c_{i} = \alpha p_{1,i} + (1-\alpha) p_{2,i} \quad \forall i=1, \ldots, \ell.\]

\subsection{Local Search}\label{local-search}

In a second step, a local search algorithm is applied to the best
solution of the genetic algorithm. Thereby, within each iteration of the
algorithm each asset weight of the $n$ assets of the portfolio is
increased or decreased by a factor $\varepsilon$. Each of these
$(2 \times n)$ new portfolios is normalized and if one exhibits a lower
fitness value then this new portfolio will be used subsequently. The
algorithm terminates if no local improvement is possible anymore or the
maximum number of iterations has been reached.

\section{Numerical Results}\label{numerical-results}

In this section the above described algorithm will be applied to
real-world financial data to obtail numerical results, which can be used
for practical portfolio optimization purposes. The first test using
stock data from the DJIA index is described in Section
\ref{financial-data-and-setup} and both the long-only case (Section
\ref{computing-djia-long-only-portfolios}) as well as the long-short
case (Section \ref{computing-djia-long-short-portfolios}) is discussed.
To check for scalability the algorithm is tested on all stocks of the
S\&P 100 index in Section \ref{scalability} afterwards.

\subsection{Financial Data and Setup}\label{financial-data-and-setup}

We use data from all stocks from the Dow Jones Industrial Average (DJIA)
index using the composition of September 20, 2013, i.e.~using the stocks
with the ticker symbols AXP, BA, CAT, CSCO, CVX, DD, DIS, GE, GS, HD,
IBM, INTC, JNJ, JPM, KO, MCD, MMM, MRK, MSFT, NKE, PFE, PG, T, TRV, UNH,
UTX, V, VZ, WMT, XOM.

Using the R package \texttt{quantmod} \cite{quantmod} we obtain daily
adjusted closing data from Yahoo! Finance. We use data from the
beginning of 2010 until the beginning of November 2014 to compute the
Variance-Covariance matrix, i.e.~the matrix is entirely based on
historical data. The data is solely used for comparison purposes such
that a clever approximation algorithm for the Variance-Covariance matrix
like those presented e.g.~by \cite{ledoit2003improved} and
\cite{ledoit2004honey} is not necessary for the purpose of this study.
However it should be noted that the matrix is the important input
parameter for the calculation.

The parameters used for the genetic algorithm are shown in Table
\ref{t:gaparam}. The local search algorithm was started twice, once with
$\varepsilon = 0.01$ and subsequently with $\varepsilon = 0.001$. The
number of maximum local search steps has been set to $500$.

\begin{table}

\centering
\caption{Parameters for the Genetic Algorithm.}

\begin{tabular}{ll}
\hline
{\bf Parameter} & {\bf Value} \\
\hline
Initial population size & 200 \\
Maximum iterations & 300 \\
\hline
Elitist selection & 10 top chromosomes from parent population \\
Random addition & 50 new chromosomes \\
Mutation & 100 chromosomes from parent population \\
Intermediate crossover & 100 pairs of chromosomes from parent population \\
\hline
\end{tabular}

\label{t:gaparam}

\end{table}

\subsection{Computing DJIA Long-Only
Portfolios}\label{computing-djia-long-only-portfolios}

First, we compute a set if various long-only portfolios without using
expected returns, i.e.~the Minimum Variance Portfolio (MVP), the 1/N
portfolio as well as the risk-parity portfolio using the algorithm
developed in this paper and described above. The results is shown in
Table \ref{t:res1}. Please note that the risk contribution has been
normalized to $1$. The fitness of the 1/N portfolio is $0.002253031$,
while the MVP exhibits a fitness of $0.00057129$. The algorithm managed
to find the Risk Parity portfolio with a fitness of $0.0005019655$. A
lower fitness is not possible due to the long-only constraint.

\begin{table}
\centering

\caption{DJIA - Long Only - MVP, 1/N, and Risk Parity.}

\begin{tabular}{rrrrrrr}
  \hline
 & x(MVP) & RCn(MVP) & x(1/N) & RCn(1/N) & x(RP) & RCn(RP) \\ 
  \hline
AXP & 0.0000 & 0.0408 & 0.0300 & 0.0444 & 0.0000 & 0.0404 \\ 
  BA & 0.0000 & 0.0374 & 0.0300 & 0.0411 & 0.0000 & 0.0366 \\ 
  CAT & 0.0000 & 0.0420 & 0.0300 & 0.0484 & 0.0000 & 0.0413 \\ 
  CSCO & 0.0000 & 0.0338 & 0.0300 & 0.0382 & 0.0000 & 0.0329 \\ 
  CVX & 0.0000 & 0.0345 & 0.0300 & 0.0357 & 0.0000 & 0.0341 \\ 
  DD & 0.0000 & 0.0382 & 0.0300 & 0.0410 & 0.0000 & 0.0376 \\ 
  DIS & 0.0000 & 0.0383 & 0.0300 & 0.0394 & 0.0000 & 0.0384 \\ 
  GE & 0.0000 & 0.0395 & 0.0300 & 0.0416 & 0.0000 & 0.0395 \\ 
  GS & 0.0000 & 0.0370 & 0.0300 & 0.0451 & 0.0000 & 0.0356 \\ 
  HD & 0.0000 & 0.0323 & 0.0300 & 0.0319 & 0.0000 & 0.0328 \\ 
  IBM & 0.0207 & 0.0285 & 0.0300 & 0.0283 & 0.0000 & 0.0272 \\ 
  INTC & 0.0000 & 0.0312 & 0.0300 & 0.0353 & 0.0000 & 0.0305 \\ 
  JNJ & 0.2015 & 0.0285 & 0.0300 & 0.0218 & 0.0376 & 0.0257 \\ 
  JPM & 0.0000 & 0.0424 & 0.0300 & 0.0502 & 0.0000 & 0.0417 \\ 
  KO & 0.0038 & 0.0285 & 0.0300 & 0.0255 & 0.0275 & 0.0334 \\ 
  MCD & 0.2421 & 0.0285 & 0.0300 & 0.0195 & 0.2333 & 0.0288 \\ 
  MMM & 0.0000 & 0.0345 & 0.0300 & 0.0359 & 0.0000 & 0.0340 \\ 
  MRK & 0.0000 & 0.0301 & 0.0300 & 0.0274 & 0.0000 & 0.0299 \\ 
  MSFT & 0.0000 & 0.0308 & 0.0300 & 0.0327 & 0.0000 & 0.0307 \\ 
  NKE & 0.0000 & 0.0343 & 0.0300 & 0.0365 & 0.0000 & 0.0347 \\ 
  PFE & 0.0000 & 0.0306 & 0.0300 & 0.0289 & 0.0000 & 0.0300 \\ 
  PG & 0.1890 & 0.0285 & 0.0300 & 0.0187 & 0.3050 & 0.0322 \\ 
  T & 0.0745 & 0.0285 & 0.0300 & 0.0228 & 0.0330 & 0.0288 \\ 
  TRV & 0.0000 & 0.0317 & 0.0300 & 0.0308 & 0.0000 & 0.0322 \\ 
  UNH & 0.0000 & 0.0305 & 0.0300 & 0.0324 & 0.0000 & 0.0293 \\ 
  UTX & 0.0000 & 0.0364 & 0.0300 & 0.0382 & 0.0000 & 0.0361 \\ 
  V & 0.0000 & 0.0330 & 0.0300 & 0.0360 & 0.0000 & 0.0320 \\ 
  VZ & 0.0554 & 0.0285 & 0.0300 & 0.0222 & 0.1072 & 0.0304 \\ 
  WMT & 0.2130 & 0.0285 & 0.0300 & 0.0176 & 0.2565 & 0.0312 \\ 
  XOM & 0.0000 & 0.0325 & 0.0300 & 0.0326 & 0.0000 & 0.0323 \\ 
   \hline
\end{tabular}

\label{t:res1}

\end{table}

\begin{figure}

\centering

\begin{tabular}{cc}

\includegraphics[width=5cm]{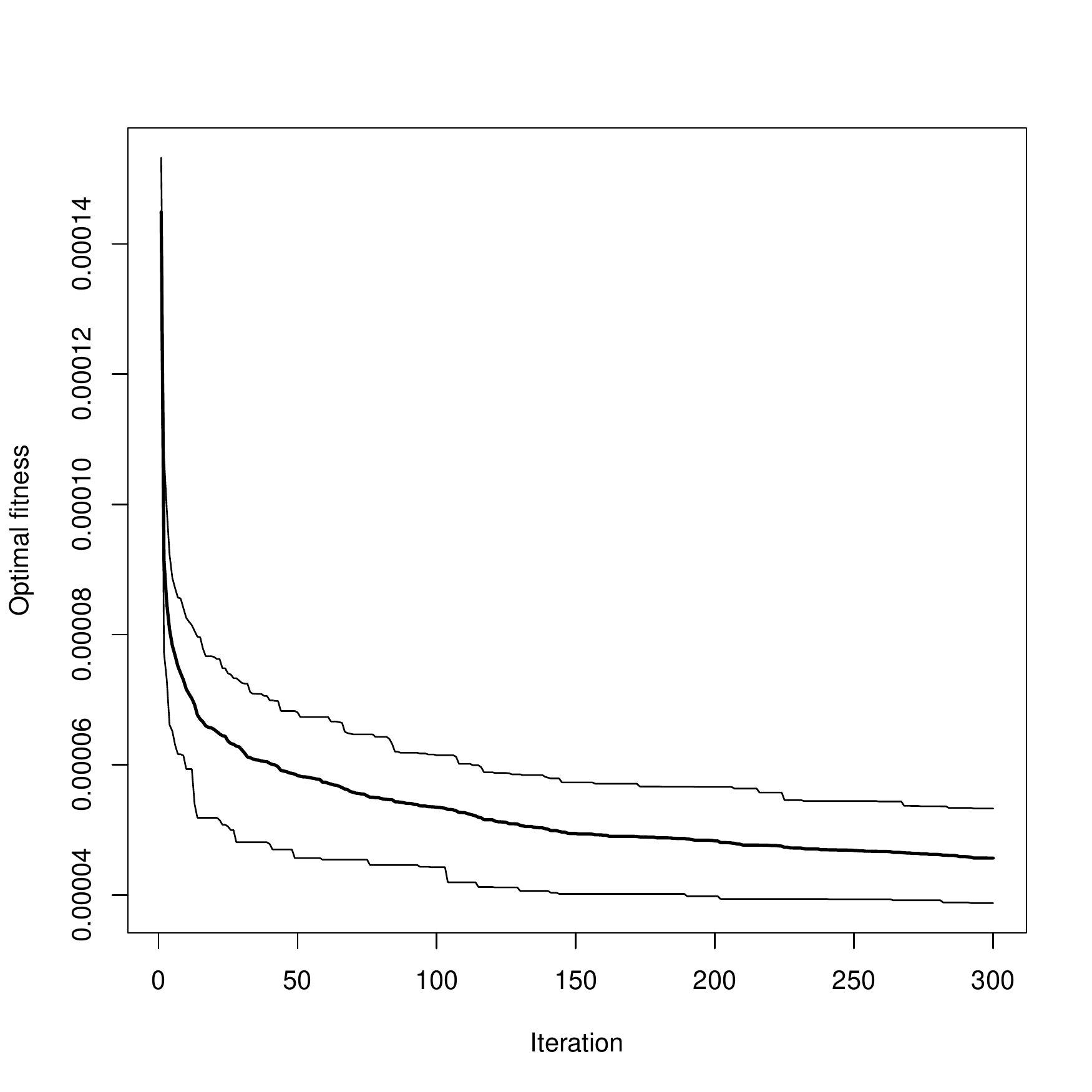} & \includegraphics[width=5cm]{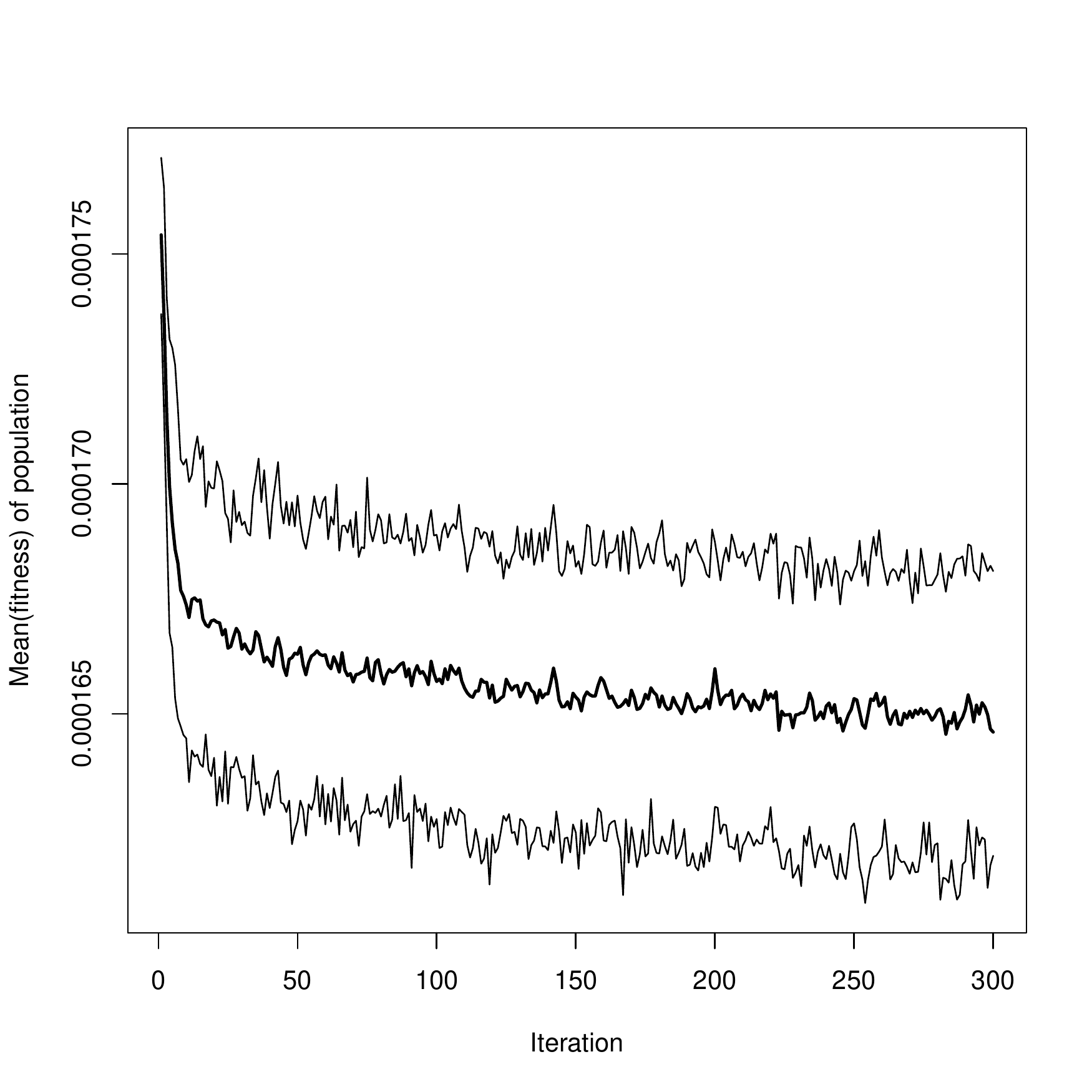} \\

\end{tabular}

\caption{Convergence of the genetic algorithm in the long-only case, i.e. the best (left) and the mean (right) fitness value of each iteration along with the 5\% as well as the 95\% quantile of 100 instances.}
\label{f:conv1}
\end{figure}

Furthermore, the convergence results in the long-only case can be seen
in Fig. \ref{f:conv1}. The left picture shows the best fitness over 300
iterations, while the right picture shows the mean of the population
fitness. The middle line depicts the mean of 100 instances while the
upper and the lower line depict the 5\% as well as the 95\% quantile of
the instances.

In the long-only case, a simple random multi-start local search
algorithm like the one described in Section \ref{local-search} above
leads to the same result. We tested this by running it $100$ times and
figured out that both the GA+Local as well as the Random+Local approach
led to the same optimal portfolio in all cases. However, the optimal
solution of the genetic algorithm needed significantly less iterations
compared to starting from random solutions. A statistical t-test
returned $t = -60.5674$ $(df = 183.198)$ and a p-value of $0$ with
respect to the number of local search iterations. However, this is
different in the long-short case, which is described in the next
section.

\subsection{Computing DJIA Long-Short
Portfolios}\label{computing-djia-long-short-portfolios}

\begin{figure}

\centering

\begin{tabular}{cc}

\includegraphics[width=5cm]{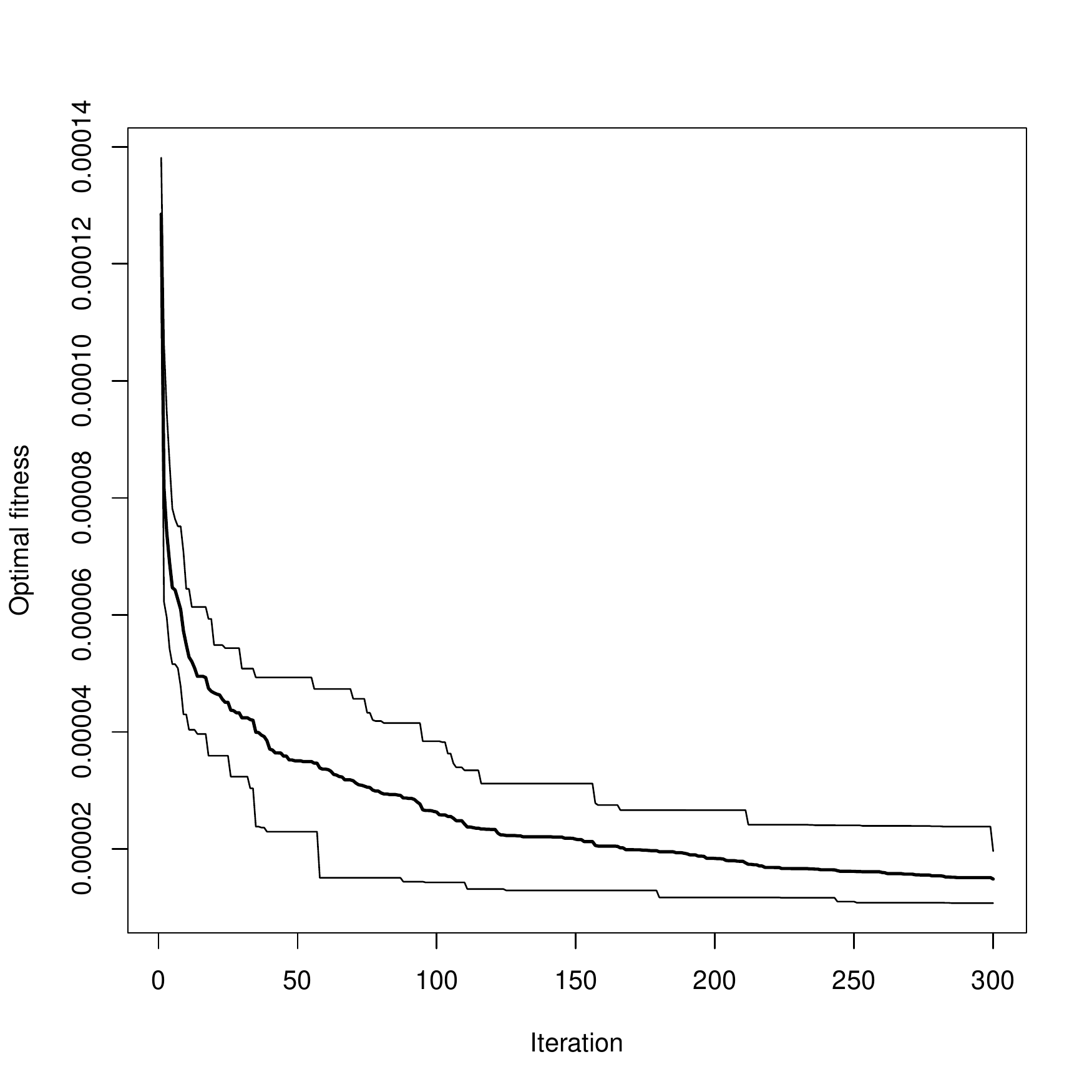} & \includegraphics[width=5cm]{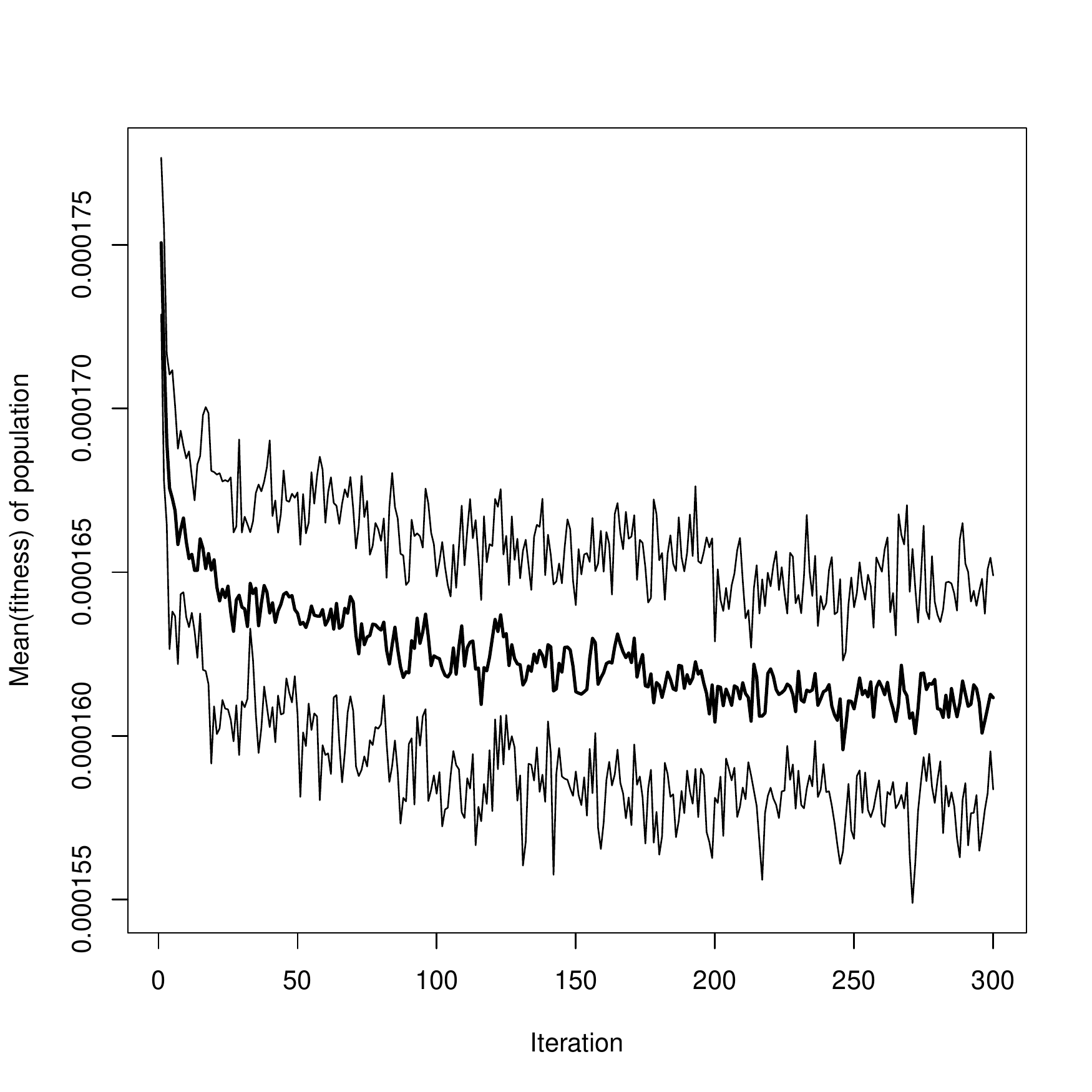} \\

\end{tabular}

\caption{Convergence of the genetic algorithm in the long-short case, i.e. the best (left) and the mean (right) fitness value of each iteration along with the 5\% as well as the 95\% quantile of 100 instances.}
\label{f:conv2}
\end{figure}

In the long-short case, a random multi-start local search heuristic does
not return any useful result. However, the evolutionary approach works
well. The long-short result with a lower bound of $-0.2$ is shown in
Table \ref{t:res2}. The convergence results in the long-short case can
be seen in Fig. \ref{f:conv2}.

\begin{table}
\centering

\caption{DJIA - Long-Short - Risk Parity.}

\begin{tabular}{rrrrrrrrrrr}
  \hline
 & AXP & BA & CAT & CSCO & CVX & DD & DIS & GE & GS & HD \\ 
  \hline
x & -0.065 & -0.010 & -0.039 & 0.000 & -0.015 & -0.042 & -0.060 & -0.071 & 0.034 & 0.019 \\ 
  RCn & 0.033 & 0.033 & 0.033 & 0.033 & 0.033 & 0.033 & 0.033 & 0.033 & 0.033 & 0.033 \\ 
  \hline
 & IBM & INTC & JNJ & JPM & KO & MCD & MMM & MRK & MSFT & NKE \\ 
  \hline
x & 0.073 & 0.024 & 0.247 & -0.050 & 0.010 & 0.257 & 0.015 & 0.012 & 0.021 & -0.004 \\ 
  RCn & 0.033 & 0.033 & 0.033 & 0.033 & 0.033 & 0.033 & 0.033 & 0.033 & 0.033 & 0.033 \\ 
  \hline
 & PFE & PG & T & TRV & UNH & UTX & V & VZ & WMT & XOM \\ 
  \hline
x & 0.016 & 0.185 & 0.102 & 0.027 & 0.016 & -0.020 & 0.013 & 0.076 & 0.211 & 0.019 \\ 
  RCn & 0.033 & 0.033 & 0.033 & 0.033 & 0.033 & 0.033 & 0.033 & 0.033 & 0.033 & 0.033 \\ 
   \hline
\end{tabular}

\label{t:res2}

\end{table}

\subsection{Scalability}\label{scalability}

To test for scalability of the algorithm, we used stocks from the S\&P
100 index as of March 21, 2014. Again, we use historical data from the
beginning of 2010 until the beginning of November 2014 to compute our
Variance-Covariance matrix. Four stocks have been excluded due to data
issues, i.e.~ABBV, FB, GM, and GOOG, such that the stocks with the
following ticker symbols have been considered: AAPL, ABT, ACN, AIG, ALL,
AMGN, AMZN, APA, APC, AXP, BA, BAC, BAX, BIIB, BK, BMY, BRK.B, C, CAT,
CL, CMCSA, COF, COP, COST, CSCO, CVS, CVX, DD, DIS, DOW, DVN, EBAY, EMC,
EMR, EXC, F, FCX, FDX, FOXA, GD, GE, GILD, GS, HAL, HD, HON, HPQ, IBM,
INTC, JNJ, JPM, KO, LLY, LMT, LOW, MA, MCD, MDLZ, MDT, MET, MMM, MO,
MON, MRK, MS, MSFT, NKE, NOV, NSC, ORCL, OXY, PEP, PFE, PG, PM, QCOM,
RTN, SBUX, SLB, SO, SPG, T, TGT, TWX, TXN, UNH, UNP, UPS, USB, UTX, V,
VZ, WAG, WFC, WMT, XOM.

\begin{figure}

\centering

\begin{tabular}{cc}

\includegraphics[width=5cm]{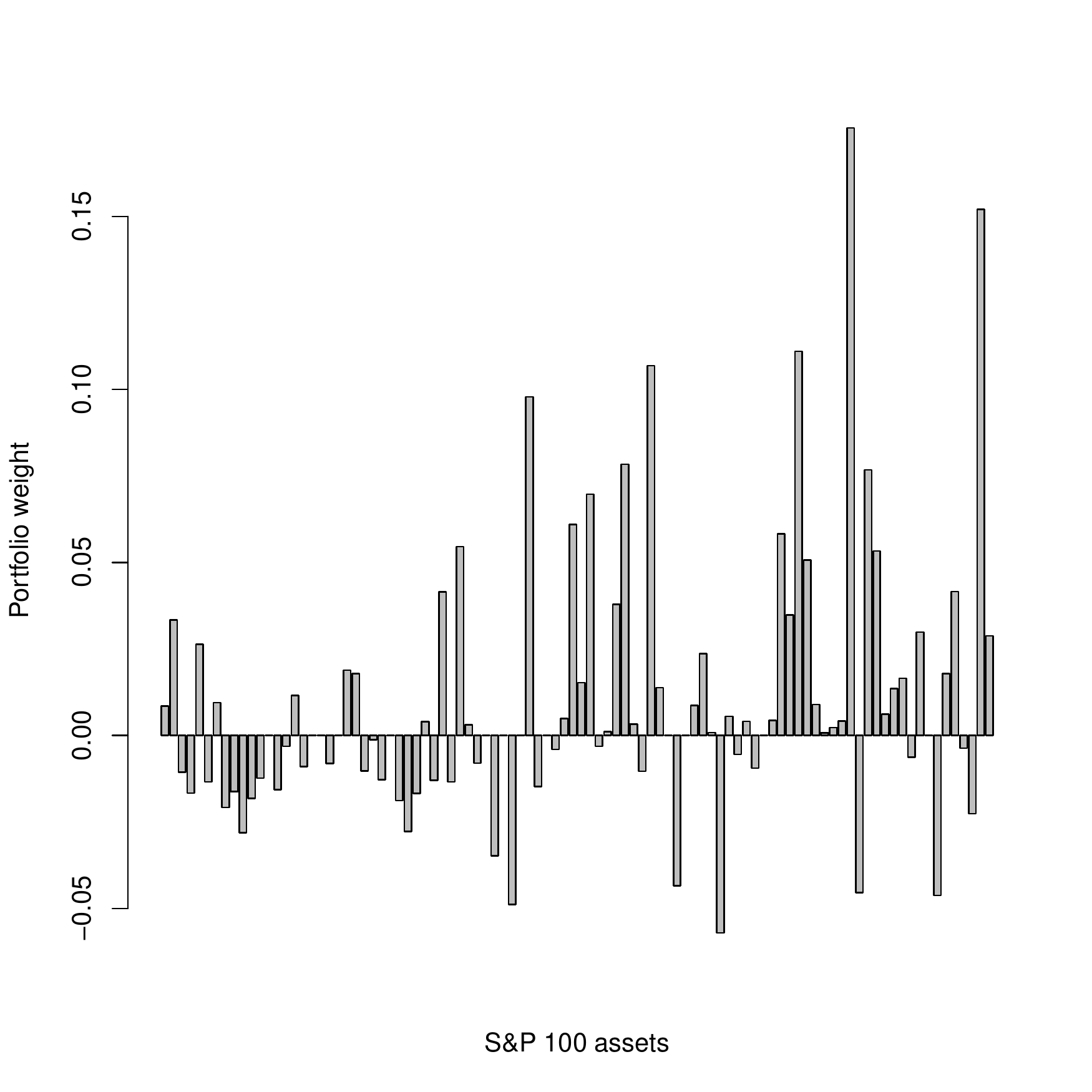} & \includegraphics[width=5cm]{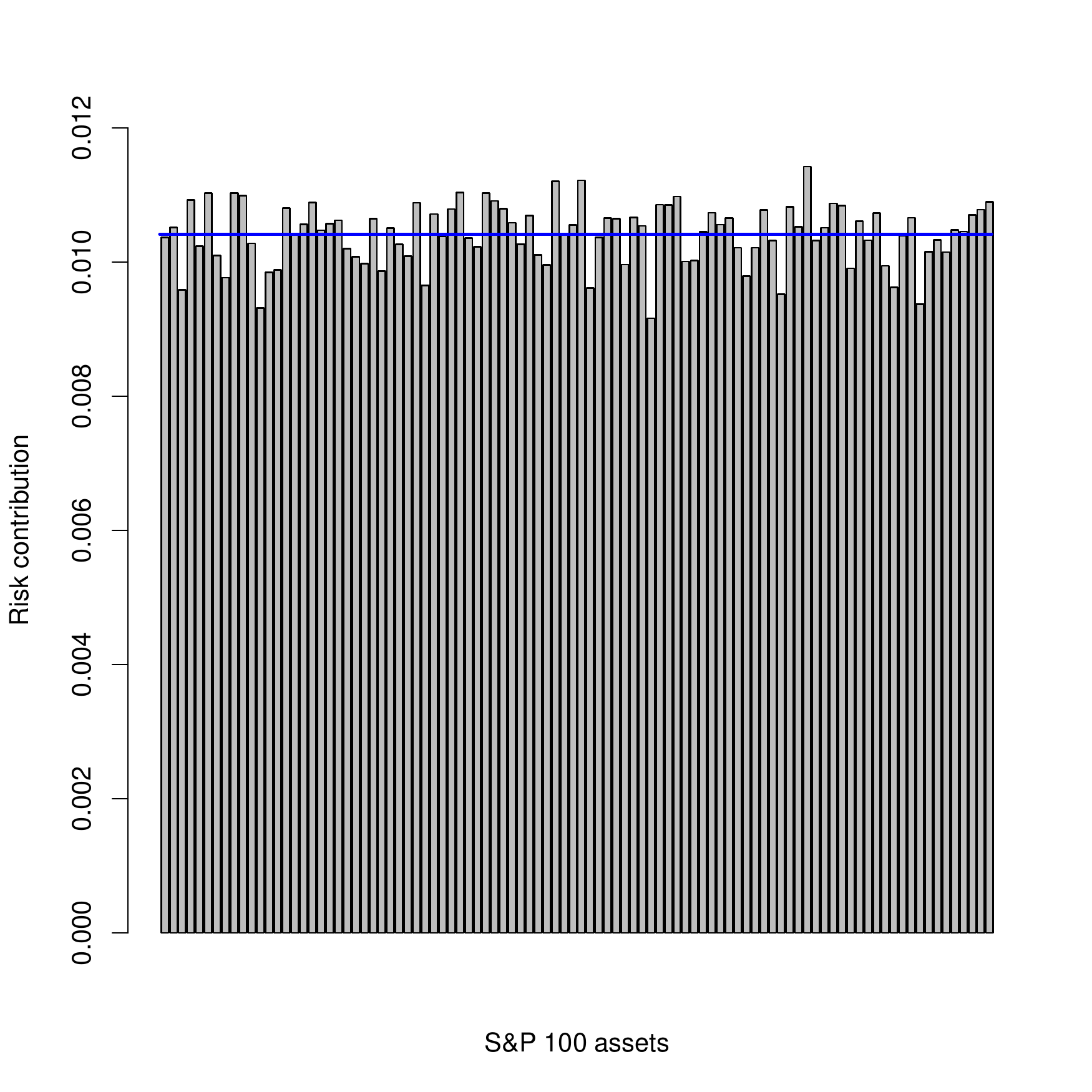} \\

\end{tabular}

\caption{S\&P 100 - portfolio (left) and risk contribution (right).}
\label{f:sp100}
\end{figure}

The lower bound was set to $-0.2$. Fig. \ref{f:sp100} shows the
resulting portfolio as well as the risk contribution of the assets. It
can be seen that the algorithm arrives at a solution, which exhibits a
rather exact risk parity solution with only slight differences from a
perfect solution, which can be observed in the right plot of Fig.
\ref{f:sp100}. To get a more detailed picture on the scalability, a
clearer analysis of the proportion between the contribution of the
evolutionary solution as well as the local search to the final solution
would have to be accomplished, but this will be left out for future
research. From an investor's perspective the optimal portfolio solution
exhibits quite a few number of assets, which would have to be shorted.
To make the solution more realistic at least a net exposure constraint
would have to be added. A cardinality constraint on the number of
shorted assets would also be an option. Both constraints can be
integrated rather easily in the evolutionary context, see e.g.
\cite{streichert2004evolutionary}, \cite{streichert2004comparing}, and
\cite{streichert2004evaluating}. However, such constraints would disable
the possiblity to obtain a perfect risk parity solution, which was the
aim of the algorithm presented in this paper.

\section{Conclusion}\label{conclusion}

In this paper, we presented an evolutionary approach to compute optimal
risk parity portfolios. This algorithm was designed to overcome the
problem that only the long-only case can be solved conveniently using
convex optimization models. A two-step approach using a genetic
algorithm as well as a local search technique proved to be successful,
especially in the long-short case. Another advantage is that further
constraints can be integrated directly into the algorithm and this
approach can be extended to other risk measures as well.

\bibliographystyle{plainnat}
\bibliography{evoparity}

\end{document}